\documentclass[lettersize,journal]{IEEEtran}
\usepackage{amsmath,amsfonts}
\usepackage{array}
\usepackage[caption=false,font=normalsize,labelfont=sf,textfont=sf]{subfig}
\usepackage{textcomp}
\usepackage{stfloats}
\usepackage{url}
\usepackage{verbatim}
\usepackage{graphicx}
\usepackage{algorithm}
\usepackage{algorithmic}
\def\BibTeX{{\rm B\kern-.05em{\sc i\kern-.025em b}\kern-.08em
    T\kern-.1667em\lower.7ex\hbox{E}\kern-.125emX}}
\usepackage{balance}
\begin{document}
\title{Compressed Geometric Arrays for Point Cloud Processing}
\author{Hoda Roodak, Mahdi Nazm Bojnordi
\thanks{Hoda Roodaki is with K. N. Toosi University of Technologys, Tehran, Iran (email: hroodaki@kntu.ac.ir)

	   Mahdi Nazm Bojnordi is with School of Computing, University of Utah, Utah, United States (email:bojnordi@cs.utah.edu)}}

\maketitle

\begin{abstract}
The ever-increasing demand for 3D modeling in the emerging immersive applications has made point clouds an essential class of data for 3D image and video processing. Tree-based structures are commonly used for representing point clouds where pointers are used to realize the connection between nodes. Tree-based structures significantly suffer from irregular access patterns for large point clouds. Memory access indirection in such structures is disruptive to bandwidth efficiency and performance. In this paper, we propose a point cloud representation format based on compressed geometric arrays (CGA). Then, we examine new methods for point cloud processing based on CGA. The proposed format enables a higher bandwidth efficiency via eliminating memory access indirections (i.e., pointer chasing at the nodes of tree) thereby improving the efficiency of point cloud processing. Our experimental results show that using CGA for point cloud operations achieves 1328$\times$ speed up, 1321$\times$ better bandwidth utilization, and 54\% reduction in the volume of transferred data as compared to the state-of-the-art tree-based format from point cloud library (PCL).
\end{abstract}


\section{Introduction}
\label{intro}
Point cloud is a set of geometric points with $x$, $y$, and $z$ coordinates that are sampled from the three-dimensional (3D) space.  Point cloud data is usually generated by computer graphics or acquired by light detection and ranging (LiDAR) scanners to represent 3D objects for various applications--e.g., medical imaging, architecture, 3D printing, manufacturing, 3D gaming, and virtual reality (VR). Point clouds store detailed information about the physical space that may lead to generating datasets in excess of terabytes. As a result, memory bandwidth and capacity become critical to high performance point cloud processing.\\
Organizing points in a data structure that supports fast point operations such as insertion, deletion, and search is a must for efficient point cloud processing \cite{Dynamic_3D_Processing}. Tree and grid are two frequently used data types for representing points \cite{b2} and \cite{3_d_cuboids}. Existing methods exploit either type in various tasks including storing, analyzing, and visualizing point clouds. A tree structure follows a node-based representation where each node implements a relationship within a subset of points. In a grid, the point cloud data is projected onto a graphical data structure to form a network of 2D images or 3D voxels \cite{PC-Processing-Framework}. \\
Tree is the most commonly used data structure for point cloud representation. For example, the point cloud library (PCL) is an open project for point cloud processing based on k-dimensional tree (kdtree) and octree \cite{Rusu_ICRA2011_PCL}. Pointers are required to represent the hierarchical connections among the tree nodes, one for each of the children. The large number of pointers used in tree results in a heavy load of memory indirections that make the point cloud processing slow.\\
As an alternative to tree-based structures, we propose compressed geometric arrays (CGA) that enable fast lookup, small memory footprint, and efficient memory bandwidth utilization. This paper is an extension of our recent work on geometric arrays for point cloud processing \cite{G-Arrays-ICASSP2021}. Here, we examine new methods for point cloud processing based on CGA.
First, we provide an introduction to various point operations such as cloud merge, projection, nearest neighbor (NN) search, and point cloud compression. Then, we investigate the challenges and opportunities of using octree in point cloud operations. Next, we explain the CGA representation format and its application to the point operations. Finally, we perform a set of experiments on various point clouds generated by LiDAR and computer graphics tools. Our experimental results indicate that CGA can improve the execution time and bandwidth utilization significantly. As compared to the PCL library, CGA achieves more than 1000$\times$ speedup for the merge, projection, and NN search operations. When used for spatio-temporal compression in MPEG G-PCC, CGA improves both the quality and compression ratio by 13-30\%.

\section{Basic Point Cloud Operations}
\label{BG & Mot}
In this section, we provide the necessary background in point cloud processing and compression.
\subsection{Merging Point Clouds}
Point cloud merging is widely used for 3D surface reconstruction in computer vision, computer graphics, and reverse engineering \cite{3D_Surface_Rec}. 
Merging structured point clouds is different from appending points of one point cloud to another. Iterative tree or graph traversals are necessary to merge point clouds, which often demand significant memory bandwidth. Point cloud merging is usually used when several views of the same object or scene are collected from different angles and positions. As each view may have a different coordinate system, a conversion of the views to a common coordinate system is necessary \cite{PCMerge}. Typically, the views are obtained from multiple 3D scanners or a single scanner positioned at different locations with specific orientations. Each view is represented as a point cloud that may have overlapping areas with other views. 
Merging point clouds consists of several stages. First, the 3D shape of an object is acquired and stored in multiple data sets. Then, the data sets are merged to a common coordinate system according to their acquisition directions  \cite{3D_Surface_Rec} \cite{3D-surface}. For merging, we consider one point cloud as the base and identify the overlapping points between the base and other point clouds according to their $x$, $y$, and $z$ coordinates. For the overlapping points, we either choose the attributes of the base, or an average of the attributes. Given two point sets $\mathbf{S_{0}}$ and $\mathbf{S_{1}}$, the following equation denotes a merge operation.

\begin{equation}
\label{eq0}
\mathbf{S_{0}} \cup \mathbf{S_{1}} = \left \{\mathbf{s_i}|\mathbf{s_i} \in \mathbf{S_{0}} \vee \mathbf{s_i} \in \mathbf{S_{1}} \right\}
\end{equation} 

\begin{figure}
  \includegraphics[width=0.5 \textwidth]{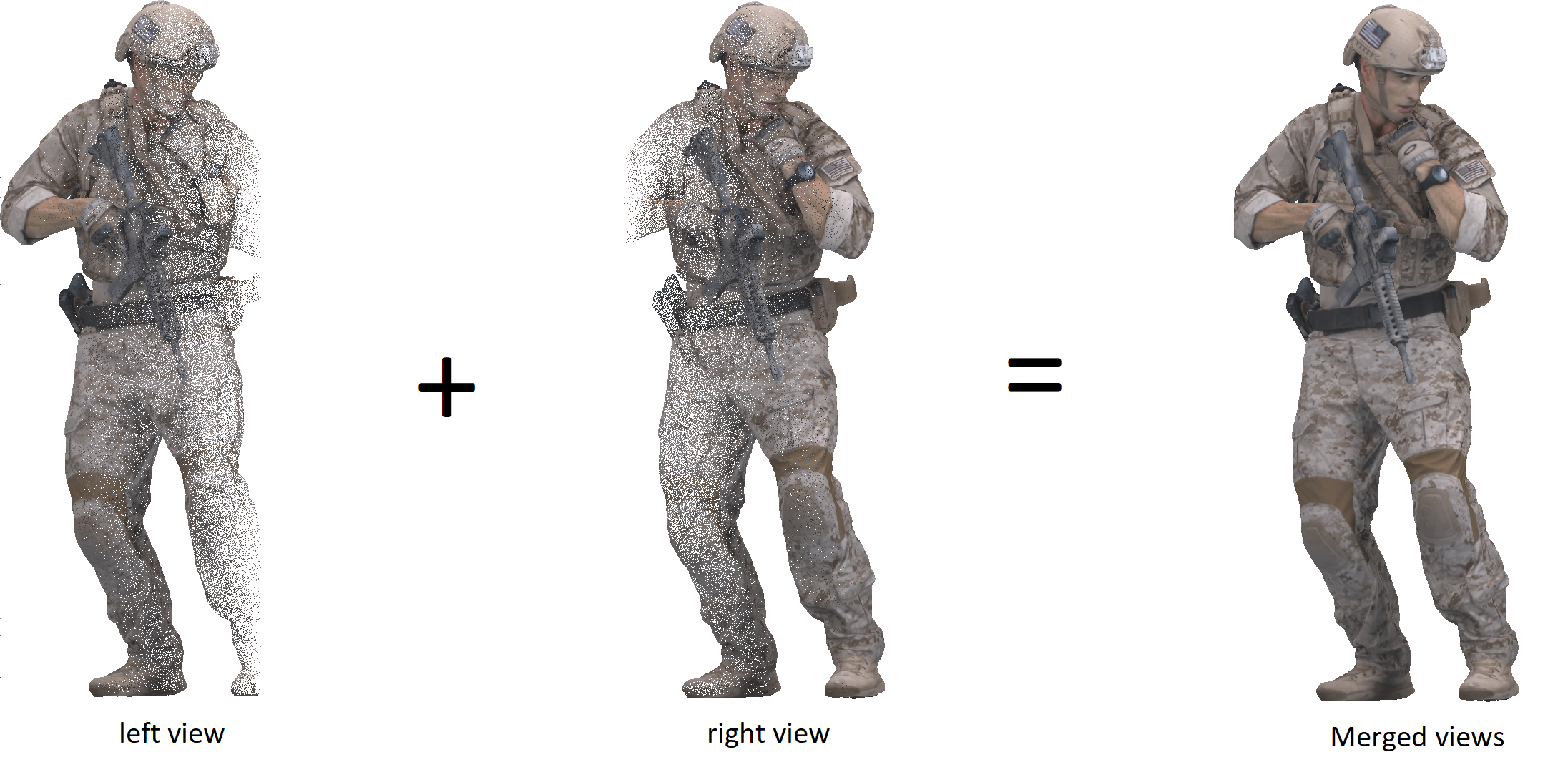}
\caption{Point cloud merging example.}
\label{merging}      
\end{figure}

In this equation, $\mathbf{s_i}$ represents a point in the 3D space with the $s^x_i$,  $s^y_i$, and  $s^z_i$ coordinates. Fig.~\ref{merging} illustrates the result of merging various views of an example point cloud frame, called Soldier \cite{CGseq}.

\subsection{3D Surface Reconstruction}
For 3D surface reconstruction, different views are first ported to a common coordinate system. Then, the initial rotation and transformation of each view are computed to find an optimal alignment among the views. For this purpose, at least three pairs of corresponding points from every two point clouds are needed to calculate a transformation between the views. For example, the following Euclidean metric $(\lambda)$ may be used to measure the similarity of  every two points in space \cite {PCmerge2}.

\begin{equation}
\begin{aligned}
\label{eq1}
\begin{split}
\lambda(\mathbf{s_0},\mathbf{s_1}) =\sqrt{\frac{(s^x_0-s^x_1)^2+(s^y_0-s^y_1)^2+(s^z_0-s^z_1)^2}{(s^x_0)^2+(s^x_1)^2+(s^y_0)^2+(s^y_1)^2+(s^z_0)^2+(s^z_1)^2}}
\end{split}
\end{aligned}
\end{equation} 

In this equation, $s_{0}$ and $s_{1}$ are two points from the same or different point clouds. A threshold similarity assessment is often performed to determine the closest points based on $\lambda$. Similar to point merging, 3D surface reconstruction requires a significant memory bandwidth. 

\subsection{Point Cloud Projection}
\label{Pointcloudprojection}
Many point cloud processing applications, such as object detecting and tracking in autonomous vehicles \cite{projIntro}, require pre-processing steps that involve projecting point clouds to 2D images. Unlike the 3D point clouds, 2D images follow a regular pattern for organizing pixels in a dense matrix. Such regularity was proved beneficial to extracting useful 2D perceptual information from 3D data that can facilitate object detection, classification, and convolutional neural networks (CNNs) \cite{projIntro}. Projection from the 3D space to 2D is usually an \textit{orthogonal} or \textit{perspective} mapping.  
\subsubsection{Orthogonal Projection}
\begin{figure}[!t]
\centering
  \includegraphics[width=0.2\textwidth]{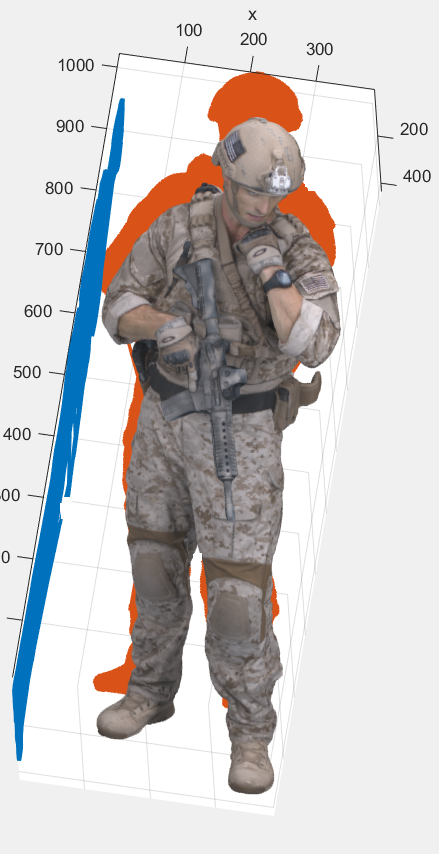}
	\caption{Orthogonal projection from 3D to 2D, 0 degrees (orange) and 90 degrees (blue).}
\label{3D2Dproj}      
\end{figure}

Orthogonal projection is generally used for representing a 3D object with three or more 2D views. In each view, the object is viewed along parallel lines that are perpendicular to the plane of that view. For example, a typical orthographic projection of a house consists of a top, a front, and a side view.
Fig. \ref{3D2Dproj} shows the orthogonal projections of an example point cloud frame from different directions. For example, consider a point $\mathbf{s} = (s^x, s^y, s^z)$ in the 3D space projected onto a 2D point $\mathbf{t} = (t^x, t^y)$ along the $z$ axis. The coordinates of $\mathbf{t}$ can be calculated as follows \cite{AutomaticObjectRecognition}.

\begin{equation}
\begin{bmatrix}
t^x \\
t^y
\end{bmatrix}
= 
\begin{bmatrix}
m_{x} & 0 & 0 \\
0 & m_{y} & 0
\end{bmatrix}
\begin{bmatrix}
s^x \\
s^y \\
s^z \\
\end{bmatrix}
\end{equation}

In this example, $m_{x}$ and $m_{y}$ are used to set the projection angle from the laser scanning direction. For instance, $m_{x} = m_{y} = 1$ is used to set the projection angle from the laser scanning direction to 90 degrees.

\subsubsection{Perspective Projection}
Perspective projection linearly maps the 3D objects onto a 2D view such that the distant objects appear smaller than the nearer ones. In a typical perspective projection, the distances, angles, and parallelism are not preserved \cite{prespective_proj}. For example, the perspective projection of a 3D point $\mathbf{s_{0}}$  with coordinates $(s^x,s^y,s^z)$ to a 2D point $\mathbf{s^\prime_{0}}$ with coordinates $(s^{\prime x}, s^{\prime y}, s^{\prime z})$  is computed by the following equation.

\begin{equation}
\begin{split}
\begin{bmatrix}
s^{\prime x} \\
s^{\prime y} \\
s^{\prime z} \\
1
\end{bmatrix}
= 
\begin{bmatrix}
1 & 0 & 0 & 0 \\
0 & 1 & 0 & 0 \\
0 & 0 & 1 & 0 \\
0 & 0 & \frac{1}{D} & 0 \\
\end{bmatrix}
\begin{bmatrix}
s^x \\
s^y \\
s^z \\
1 \\
\end{bmatrix}
\end{split}
\end{equation}

In this projection, the projection plane is considered at $-D$ on the $z$-axis. The 2D coordinates are computed by the following equations. 

\begin{equation}
\label{eq2}
s^{\prime x} = \frac{s^x.D}{s^z}, s^{\prime y} = \frac{s^y.D}{s^z}, s^{\prime z} = D
\end{equation}

Note that in a perspective projection, one of the $s^x$, $s^y$, or $s^z$ components may be omitted to form a 2D view \cite{prespective_proj}. Moreover, vertex processing, modeling, and viewing transformation may be required before the projection. For instance, to view 3 faces of a cuboid, one rotation is used before applying the perspective projection. For a rotation with a $\theta$ angle around the y-axis, the following matrix operations are necessary.
In all, point cloud projection requires visiting points to compute their projected coordinates. In a tree structure, such data stream is generated by traversing the tree nodes that results in irregular access patterns and low bandwidth utilization.

\begin{equation}
\begin{split}
\begin{bmatrix}
s^{\prime x} \\
s^{\prime y} \\
s^{\prime z} \\
1
\end{bmatrix}
= 
\begin{bmatrix}
cos(\theta) & 0 & -sin(\theta) & 0 \\
0 & 1 & 0 & 0 \\
sin(\theta) & 0 & cos(\theta) & 0 \\
0 & 0 & 0 & 1 \\
\end{bmatrix}
\begin{bmatrix}
1 & 0 & 0 & 0 \\
0 & 1 & 0 & 0 \\
0 & 0 & 1 & 0 \\
0 & 0 & \frac{1}{D} & 0 \\
\end{bmatrix}
\begin{bmatrix}
s^x \\
s^y \\
s^z \\
1 \\
\end{bmatrix}
\end{split}
\end{equation}

\subsection{Point Cloud NN Search}
Consider a 3D space $\mathbf{S}$ and a query point $\mathbf{s_0} \in \mathbf{S}$. An NN search refers to selecting 1 point from $\mathbf{S}$ such that it is closer to $\mathbf{s_0}$ than the rest of points in $\mathbf{S}$. Several metrics exist for measuring the distance between every two points ($\mathbf{s_0}$ and $\mathbf{s_1}$) in space. For example, Euclidean distance ($\varepsilon$) is defined by the following equation.

\begin{equation}
\label{eq5}
\begin{split}
\varepsilon(\mathbf{s_0},\mathbf{s_1}) =
\sqrt{(s^x_0-s^x_1)^2+(s^y_0-s^y_1)^2+(s^z_0-s^z_1)^2}
\end{split}
\end{equation}

Given a point set $\mathbf{S}$ with $n$ points and a query point $\mathbf{s_i}$ ($\mathbf{s_i} \in \mathbf{S}$), NN is to find a subset $\mathbf{C}$ $(\mathbf{C} \subset \mathbf{S}) $ containing nearest points to $\mathbf{s_i}$. For all $\mathbf{s_0} \in \mathbf{C}$ and $\mathbf{s_1} \in \mathbf{S}$-$\mathbf{C}$ (the set whose elements belong to $\mathbf{S}$ but not $\mathbf{C}$):

\begin{equation}
\label{eq6}
\varepsilon(\mathbf{s_i},\mathbf{s_0}) \leq \varepsilon(\mathbf{s_i},\mathbf{s_1}) 
\end{equation}

\begin{figure}[!t]
\centering
  \includegraphics[width=0.4\textwidth]{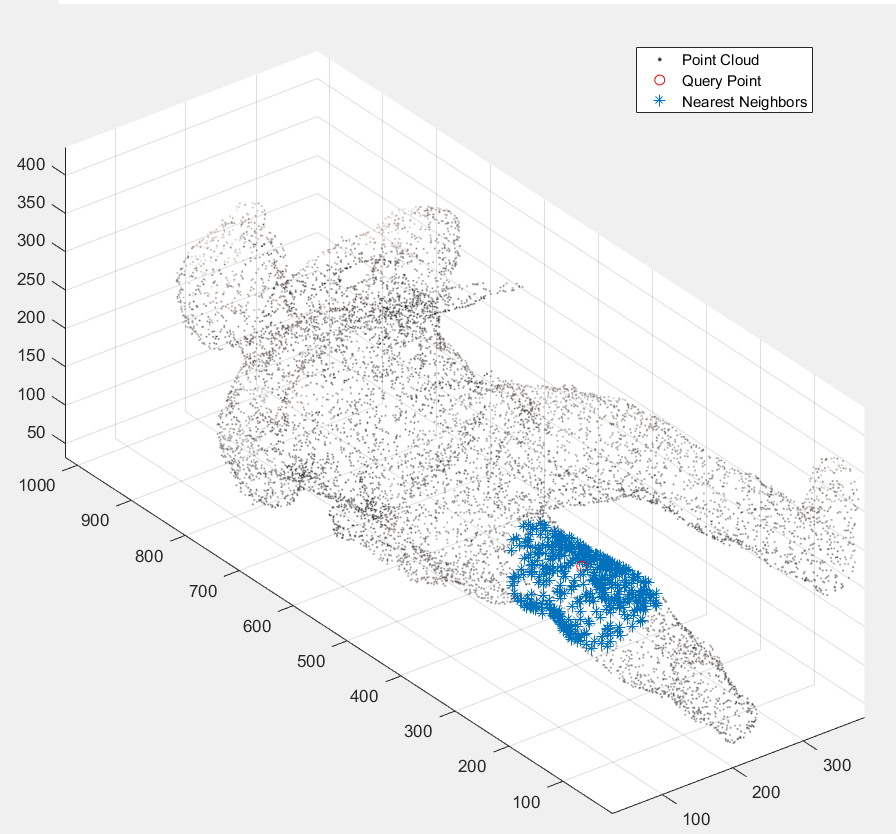}
	\caption{Point cloud 400 nearest search example.}
	\label{kNearest}
\end{figure}

Fig.~\ref{kNearest} shows the results of an example NN search to find 400 nearest point.
Most NN methods are based on recursive subdivisions of the 3D space \cite{KNN2}. For example, octree relies on uniform subdivisions \cite{KNN2} while kdtree uses a nonuniform subdivision approach \cite{KNN2}. A kdtree based NN algorithm is described in Algorithm \ref{alg:KNN}.

\begin{algorithm}
\caption{NN Search}
\begin{algorithmic} 
\label{alg:KNN}
\STATE \textbf{Input}: \\
\STATE A dataset \textbf{S}\\
\STATE A query point q\\
\STATE Tree root: the root of a Kdtree\\
\STATE A non-leaf node in Kdtree divides the space into two parts: 
\STATE \hspace{\algorithmicindent} left subtree: points to the left of this space  
\STATE \hspace{\algorithmicindent} right subtree: points to the right of the space
\STATE Superior: A node that has a child is called the child's superior
\STATE \textbf{Output}: the k closest points $s_{1},s_{2},...,s_{k} \in \textbf{S}$ to q
\FOR {each p point}
\STATE Start at the tree root
\STATE //Traverse the kdtree to the subtree the p point belongs
\STATE Find the leaf
\STATE Store it as the "current best point".
\STATE //Traverse upward 
\FOR {each superior node of the "current best point" }
\IF {the superior point is closer to the q point than the current best point}
\STATE select it as the current best point.
\ENDIF
\ENDFOR
\STATE //Check the subtree on the other side
\IF {there is a closer point to p in this subtree}
\STATE traverse this subtree to a leaf
\ENDIF
\STATE //The nearest neighbor point is found
\ENDFOR
\end{algorithmic}
\end{algorithm}

\subsection{Point Cloud Compression}
\begin{figure}[!t]
\centering
  \includegraphics[width=0.5\textwidth]{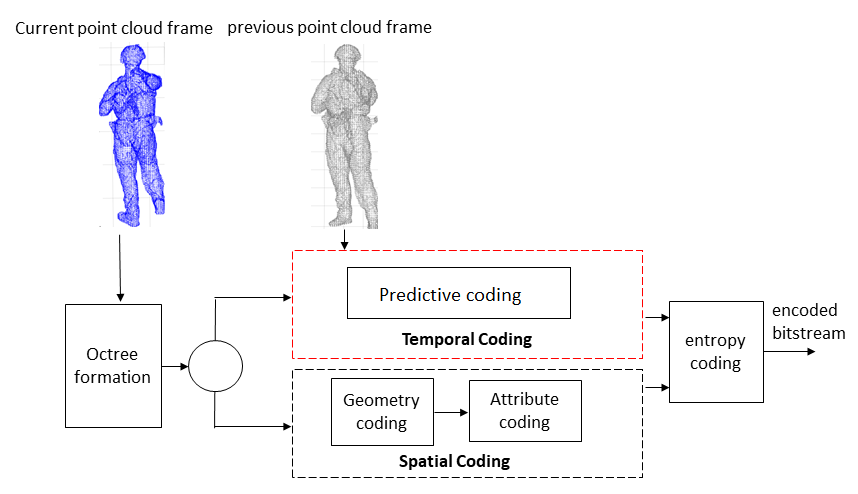}
	\caption{Schematic of point cloud compression \cite{b1}.}
	\label{G-PCC_a}
\end{figure}

Generally, point clouds are of two types: \textit{static} and \textit{streaming}. Static point clouds are suitable for representing one object, such as a building or a human face. In contrast, streaming point clouds are used to represent the progress of an event in time. For compressing such data, the redundancies among neighboring frames may be extracted to enhance the quality of compressed data \cite{b03}. Most of these methods are based on octree formation of the 3D space in which the point cloud is encoded in terms of occupied octree cells. Fig.~\ref{G-PCC_a} shows the most important components of the 3D geometric point cloud compression, combines features from common 3D octree-based point cloud compression and common hybrid video coding that proposed in traditional codecs such as H.264 and HEVC which include block based motion compensation \cite{b1}.

\subsubsection{Constructing Octree}
An octree is defined as a tree with each node comprising up to 8 children. The octree of a 3D space is created by recursively dividing it into octants.
The G-PCC encoder recursively divides the point cloud aligned bounding box into eight children. Only non-empty children continue to divide.\\
An 8-bit code is used to represent each octree subdivision. A 1-bit flag is used to denote whether a child cell is empty or not, with ‘1’ indicating a nonempty child cell and ‘0’ an empty child cell. The cells are traversed  according to a fixed order and the flag bits of all child cells are collected to obtain an 8-bit code, which is called the occupancy code as shown in Fig.~\ref{G-PCC_b}.
Starting from root node, each octree level has an 8 bit occupancy code. Only non-empty nodes are divided further. The decoder only needs the occupancy codes to reconstruct an octree.

\subsubsection{Spatial Coding}
The Spatial coding consists of three steps \cite{b1}.

\begin{figure}[!t]
\centering
  \includegraphics[width=0.4\textwidth]{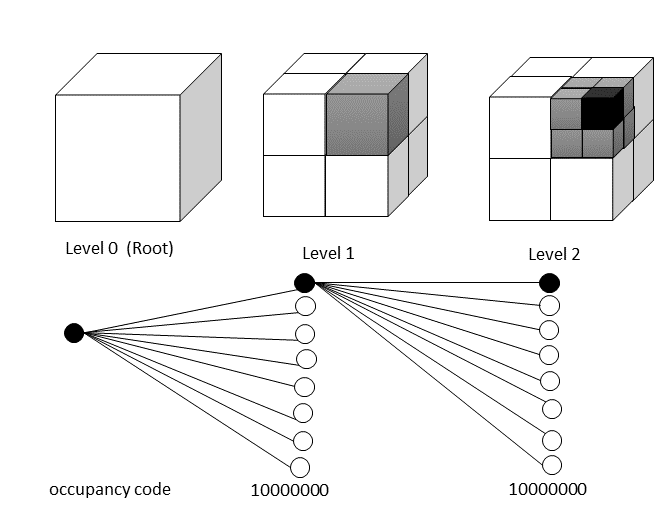}
	\caption{Illustrative example of an octree and its occupancy code based on the prior work \cite{b1}.}
	\label{G-PCC_b}
\end{figure}

\paragraph{Bounding Box Alignment}
A box with a lower corner at ($s^x_{min}$,$s^y_{min}$,$s^z_{min}$) and an upper corner at ($s^x_{max}$,$s^y_{max}$,$s^z_{max}$) is called a bounding box. This box is used as the root of the octree and may change from one frame to another. The spatio-temporal locality of point coordinates may vary between consecutive frames. Therefore, the geometric correspondence of points between consecutive frames may be lost, which makes temporal prediction hard. Bounding box alignment is a technique for adjusting the size of bounding box so to include all the points \cite{b1}.
\paragraph{Coding of the Occupancy Codes}
Starting from the root, each octree subdivision results in an 8 bit occupancy code as shown in Fig. \ref{G-PCC_b}. Since, only non-empty nodes are subdivided, the decoder only needs the occupancy codes to reconstruct the point cloud. So, efficient coding of occupancy codes is crucial.

\paragraph{Color Encoder}
In addition to the geometry information, the color attributes should be coded in point cloud compression. To encode the color attributes efficiently, the redundancy between them should be extracted. For this purpose, the octree is traversed based on the depth first search method in which, first, all descendants of root are visited recursively and then the root is visited. Then, the color attributes are written into some 8$\times$8 blocks. Since, in the depth first traversal, the consecutive pixels are often co-located and therefore are correlated to each other, the redundancy between them could be extracted using discrete cosine transform (DCT) and Differential Pulse Code Modulation (DPCM). After applying DCT, DC component is large and varied, but often close to previous value. DPCM proposes to encode the difference between the DC value of each 8$\times$8 block from the previous block. Algorithm \ref{alg:color coding} shows multiple steps of color coding \cite{b1}.

\begin{algorithm}
	\caption{Color coding in point cloud compression}
	\begin{algorithmic} 
		\label{alg:color coding}
		\STATE Start at the root of octree
		\STATE Traverse the octree based on the depth first search
		\STATE //First visits all descendants of root recursively, finally visits the root 
		\FOR {all visited points} 
		\STATE Extract the color attributes
		\ENDFOR
		\STATE All extracted color attributes are written into some 8×8 blocks. 
		\STATE // In depth first search, the consecutive pixels are often co-located and therefore are correlated to each other
		\STATE Apply DCT to 8$\times$8 blocks
		\STATE Use DPCM to encode the difference between the DC values of each 8×8 block from the previous block.
		\STATE Send the residuals of DPCM
		\STATE Send the AC values
	\end{algorithmic}
\end{algorithm}

\paragraph{Temporal Prediction}
The temporal prediction algorithm uses the spatial coding method in combination with a prediction scheme. The data in predicted frame (P-frame) is encoded in two parts. The first part is data that contains the points that could not be predictively coded and the second part is the data that could be predicted well from the previous frame.\\
For the first part, the spatial coding method is used and for the second part, the cubes with appropriate size, 8$\times$8$\times$8, 16$\times$16$\times$16 or 32$\times$32$\times$32 are generated. Then, each one is traversed to find a corresponding cube in reference frame (I-frame). The color variance of the points in the cube is also checked and the temporal prediction only performs in the area of point cloud that have low color/texture variance. The temporal prediction may result in visible artifacts in high variance texture image regions.  the prediction is performed based on computing a rigid transform between the matching cubes. The rigid transform is a $4\times4$ matrix that is composed of a $3\times3$ rotation matrix and a $3\times1$ translation vector. This computation is based on the iterative closest point (ICP) algorithm and only takes the geometric coordinates into account. If the iterative closest point algorithm converges, the predictor, which is the rigid transform and the position of $cube_{P}$ are coded. Otherwise the cube is coded using spatial coding. Using the matching cube in I-frame, the motion vector is computed. Finally, a color offset is  coded to compensate for color difference offsets between the corresponding cubes. This color offset can be used to compensate the color difference between cubes due to the brightness difference. Algorithm \ref{alg:predictive coding} show the predictive point cloud coding algorithm \cite{b1}. 

\begin{algorithm}
	\caption{Temporal point cloud coding}
	\begin{algorithmic} 
		\label{alg:predictive coding}
		\STATE Generate the cubes of P-frame and I-frame 
		\FOR {each $cubes_{P}$ in P-frame do}
		\STATE Seach to find the corresponding cube, $cube_{I}$ in I-frame
		\IF{$cube_{I}$ not found}
		\STATE use spatial coding for $cube_{P}$
		\ELSE
		\STATE check the color variance $cube_{P}$ points
		\IF {the color variance $\ge$ threshold}
		\STATE computing a rigid transform between $cubes_{P}$ and $cube_{I}$ using ICP
		\IF {ICP converged} 
		\STATE code rigid transform 
		\STATE code position of $cube_{P}$
		\ENDIF
		\ELSE
		\STATE spatial coding for $cube_{P}$ 
		\ENDIF   
		\STATE code the color offset between $cubes_{P}$ and $cube_{I}$
		\ENDIF
		\ENDFOR
	\end{algorithmic}
\end{algorithm}

\section{Design Challenges and Opportunities}
Traversing an octree suffers from costly memory indirections due to pointer-chasing at the tree nodes. Memory access indirection is counterproductive to bandwidth efficiency and performance of point cloud processing due to lack of locality in memory accesses. In the litrature, there are several implementations for octrees \cite{Octrees_representation}.
\subsection{Octree Implementation}
\subsubsection{Standard Representation}
In this representation, each tree node includes eight pointers. The leaf nodes have no children; therefore, two node types are used to form a tree. One for the inner nodes and one for leaves. Moreover, a flag must be used to distinguish between an inner and a leaf. Some octree implementations add pointers to the parent node for bottom-up traversals.
\subsubsection{Block Representation}
To save memory, it is possible for each node to store only one pointer to a block of eight children, instead of eight individual pointers. However, all eight children must be allocated when a leaf node is subdivided and it is not possible to allocate child nodes on-demand.
\subsubsection{Sibling-child Representation}
In this method, each node uses two pointers per node. One for the next sibling, which is the next child of the node’s parent. Another pointer is for the first child of the node. In comparison to the standard representation, this representation needs less memory for pointers.

\begin{figure}[H]
\centering
  \includegraphics[width=0.5\textwidth]{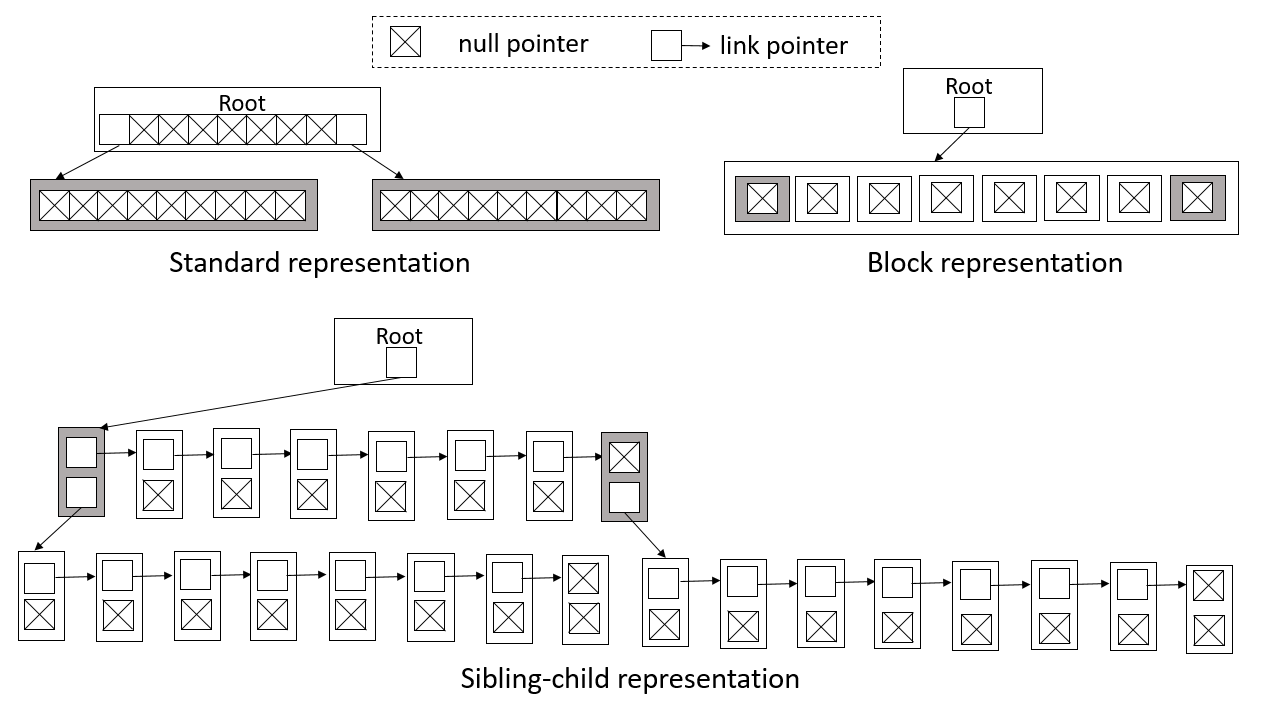}
	\caption{Various octree structures based on the prior work \cite{Octrees_representation}.}
	\label{Octrees_representation}
\end{figure}

In the standard representation, the best case is when the first child should be accessed and only one indirection is required. In the worst case, where the last child should be accessed, eight indirection is required. The block representation requires fewer indirections than the standard representation.
In sibling-child representation, two pointers instead of eight pointers are required per node; therefore, it consumes less memory than the standard representation. When the child nodes are accessed sequentially, the indirections in this representation is the same as the standard representation. But, the worst case is when the node should be accessed randomly, in which the indirections is much higher than the standard representation \cite{Octrees_representation}. Fig. \ref{Octrees_representation} shows the difference between various representation of octree.

\subsection{Performance Analysis}
\label{Performance_analysis}
\begin{figure*}[htb!]
\centering
  \includegraphics [width=0.95\textwidth, height=0.95\textheight]{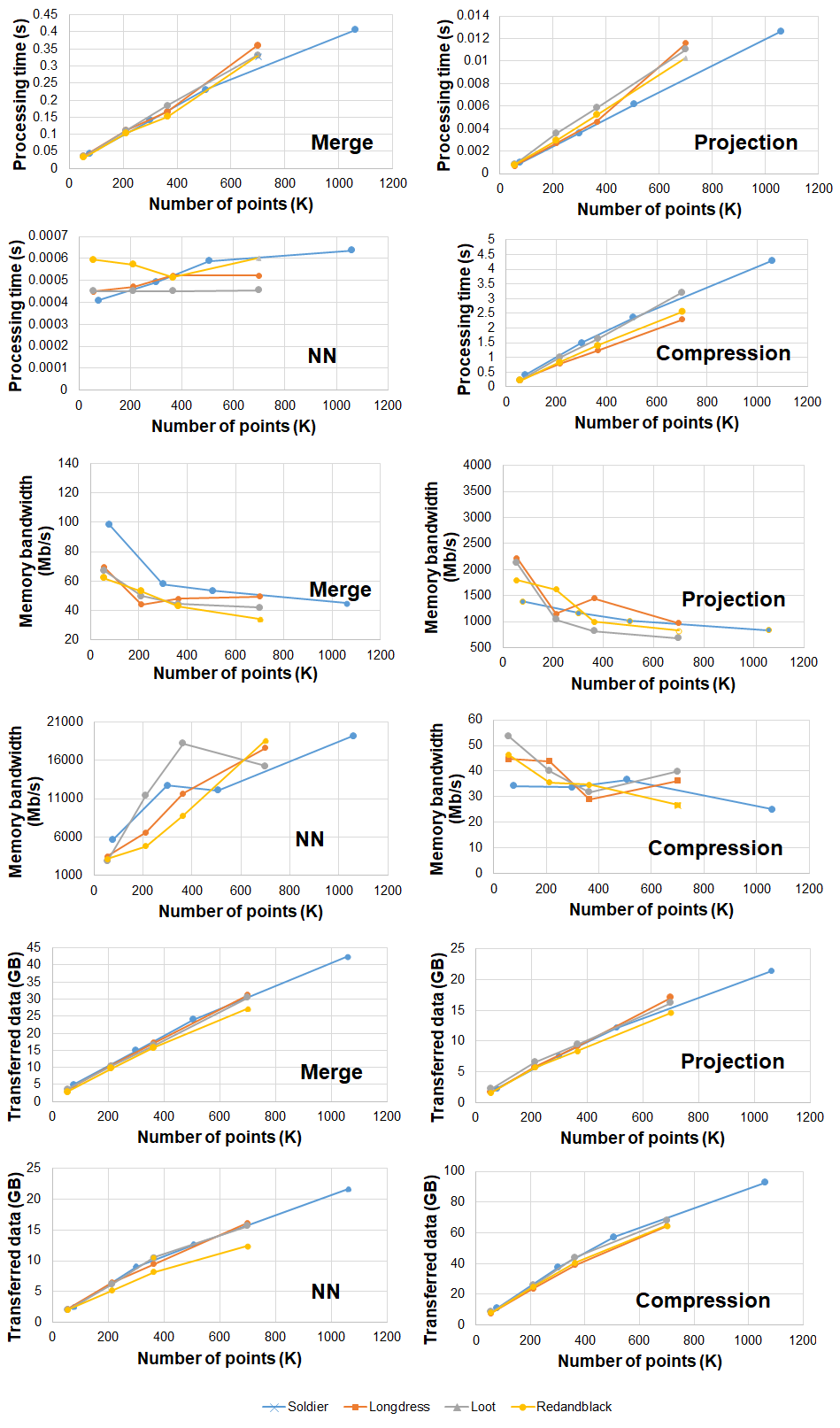}
	\caption{Processing time, bandwidth utilization, and the volume of data transfer for different point cloud processes and various computer-generated point cloud sequences using PCL library.}
	\label{PCL_PT}
\end{figure*}

\begin{figure*}[htb!]
\centering
  \includegraphics [width=0.95\textwidth,  height=0.95\textheight]{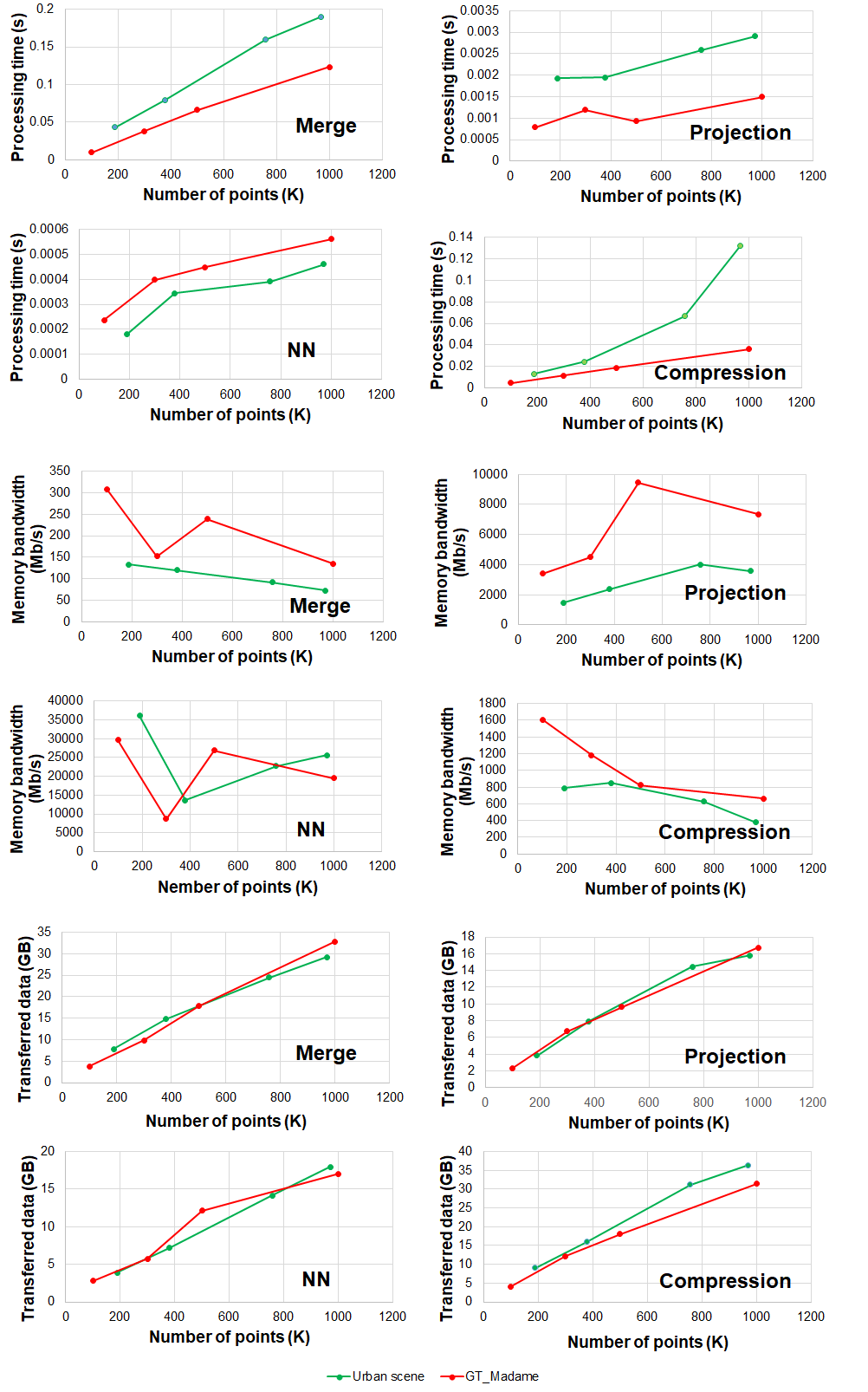}
	\caption{Processing time, bandwidth utilization, and the volume of data transfer for different point cloud processes and various LiDAR point cloud sequences using PCL library.}
	\label{PCL_PT_Lidar}
\end{figure*}

To investigate the computational complexity, memory bandwidth utilization, and the volume of data transfer, we use four different computer-generated sequences that capture human bodies in motion, i.e., Soldier, Longdress, Loot, and Redandblack \cite{CGseq} and two LiDAR sequences \cite{LiDAR_seq}. Various point cloud processes, i.e., point cloud merging, 3D to 2D projection, NN and point cloud compression are implemented on a system with Intel(R) Core(TM) i7-8750H CPU @ 2.5GHz 2.6 GHz and 8.00 GB RAM. We use the octree structure of PCL library \cite{Rusu_ICRA2011_PCL} for point cloud representation. \\
PCL is a C++ library for 3D point cloud processing, in which most mathematical operations are implemented with and based on Eigen, an open-source library for linear algebra. Considering the efficiency and performance of modern CPUs, PCL provides support for OpenMP \cite{openmp} and Intel Threading Building Blocks (TBB) library \cite{Intel_threading} for multi-core parallelization. To make the algorithms more efficient, the modules in PCL pass data around using shared pointers which avoiding the need to re-copy data presented in the system.  So, the primary data structures in PCL are fully optimized \cite{Rusu_ICRA2011_PCL}. \\
PCL includes various algorithms for point cloud processing, such as filtering, surface reconstruction, segmentation, etc \cite{Rusu_ICRA2011_PCL}. The PCL as the most famous open source library for 3D point cloud processing is released under the terms of the BSD license, which is free for commercial and research use \cite{PCL}. PCL has two space-partitioning data structures, kd-tree and octree that both of them use a tree structure to store the points. Based on the benefits listed for PCL, the octree structure of this library is used in this paper as the baseline structure. \\
The processing time, bandwidth utilization, and the volume of data transfer is measured for computer-generated and LIDAR sequences and are shown in Fig. \ref{PCL_PT} and Fig. \ref{PCL_PT_Lidar}, respectively. The x-axis shows the number of points for each sequence. We vary number  of points by downsizing the point clouds to three different versions in addition to their original sequence. For this purpose, the pcdownsample function of MATLAB is uses.  In this function, the points within a 3D box with a specific size (Gridstep) are merged to a single point in the output sequence. Gridstep is the input of pcdownsample function which is specified as a numeric value. This value should be small when there are numerous number of points in 3D space to construct fine-grained grids. Then, the color attributes of the points whiten a 3D box are averaged accordingly. This method preserves the shape of the point cloud. \\
The processing time, memory bandwidth utilization, and the amount of transferred data for different operation are then measured using the Intel VTune Profiler \cite{VTune}. Memory bandwidth utilization is defined as the LLC(last-level cache) miss counts divided by the processing time. The LLC is the last, and longest-latency, level in the memory hierarchy before main memory (DRAM). Any memory requests missing here must be serviced by local or remote DRAM, with significant latency. The LLC miss count metric shows total number of demand loads which missed LLC.  The high amount of bandwidth consumed shown in Fig. \ref{PCL_PT} and Fig. \ref{PCL_PT_Lidar}, indicate the requirement to improve the bandwidth utilization for various point cloud processes.

\section{Related Work}

\subsection{Data Structures for Point Clouds}
Numerous data structures have been proposed for storing point cloud data in the literature. Windowed priority queue is a data structure for point clouds that holds references to the first dimension of points. Then, the intervals of points are identified by the windowed priority queue. This approach focuses on the indexing of the data for fast retrieval and can be generalized for all dimensions of point data by instantiating the windowed priority queue repeatedly \cite{WinPQ}.\\
Octree is another implementation of point cloud that refers to a tree in which each node has up to eight children and represents the volume formed by a rectangular cuboid \cite{b2}. The internal nodes in octree used to represent a 3D region or a cuboid and the leaf nodes are used to to represent that no point exists in the region it represent. When storing a point cloud, the maximum depth is defined as a stopping rule for occupied volumes. \\
In octree, most information about the inner nodes is extracted using the information of neighboring nodes. For example, the depth of each node is calculated as the depth of its parent plus one and the parent pointers are pushed to a stack to remember the parents of each node. So, the information of each node that could be obtained by traversing the tree is not saved in memory and would be omitted. In addition, one byte is added to each node in which each bit corresponds to one octant of the node. So, it is not required to store 8 child for each node to make the memory consumption more efficient.\\
The data structure proposed in \cite{3_d_cuboids} extends the Voronoi diagram and Delaunay triangulation \cite{Voronoi_diagrams} to an environment for 3D voxels. But, it is not clear how we can efficiently scale this approach to handle large scale point cloud data.\\
A kdtree proposed in PCL is a space-partitioning data structure for organizing points in a k-dimensional space. It tries to partition the space for organizing points in a k-dimensional space. It splits the space into two parts in each layer. Some or half of the points may be stored in the left subtree, and the other ones in the right subtree. It stops splitting when the number of points in one node reaches the specified value \cite{Kd-tree}. The main drawback of this structure is that its performance degrades rapidly if new points are required to be added to the kd-tree after its construction \cite{Spatial-DS}.
Some implementations of octree store redundant information for each node. For instance, some pointers to neighbor and parent nodes are saved for fast tracking of nodes in search operations. In some other implementations, the position and size of each node and eight pointers to child nodes are stored. So, the subdivision could be stopped fast to reduce the required number of nodes for storage \cite{b2}. 
Serialized pointer-free octrees are the most memory efficient structure. But, the accessing time in this structure is O(n), where n is the size of the data (number of points/cells). This implementation is useful in bandwidth intensive applications \cite{b2}.

\subsection{Point Cloud Processing}
Machine learning on point clouds has been attracting more attention in recent years. Fundamental processes in 2D images, such as including 3D shape classification, 3D object detection and tracking, and 3D point cloud segmentation also exist in point clouds \cite{Deep_Learning-survey}.\\
The 3D shape classification methods are classified into multi-view and volumetric-based methods according to the input data type. In the multi-view methods, first, the 3D shape is projected to multiple views to extract the view-wise features. Then, these features are used for shape classification. Finding the view-wise features is the main challenge for these methods. The volumetric-based methods try to voxelize a point cloud into a 3D grids to apply a 3D convolution neural network for shape classification.\\
A 3D object detector takes the input point cloud of a scene and generates an 3D bounding box around each detected object. These methods can be divided into two categories: region proposal-based and single shot methods. The proposal-based methods first extract several possible regions (proposals) containing objects. Then, the region-wise
features are used to determine the category label of each region. The single shot methods try to predict various categories and then regress the predicted bounding boxes of objects using a single-stage network \cite{Deep_Learning-survey}.\\
3D object tracking uses the locations of an object in the first frame to estimate its location in subsequent frames. The geometric information in point clouds can overcome main drawbacks of object detection in 2D images such as occlusion and illumination and scale variation \cite{Deep_Learning-survey}.\\
Automated driving has been developed rapidly in recent years. It requires vision-based SLAM (simultaneous localization and mapping) systems. But, static environment is a prerequisite for these systems to work properly, which greatly limits the use of SLAM in real 3D environments \cite{Autonomous_driving} . In addition, a mono-camera perception system cannot provide reliable 3D geometry, which is essential for autonomous driving. Therefore, autonomous vehicle are usually equipped with a suite of various sensors to ensure accurate environmental perception. This way, the camera LiDAR fusion is becoming an emerging research area. Processing the data gathered from such sensors has a main challenge. In terms of data structure, the point cloud is irregular, orderless and continuous. These characteristics differences between the point cloud and the image, which is while the image is regular, ordered and discrete, make processing of point cloud a challenging task \cite{SLAM}.
Virtual and augmented reality created using point clouds is another attractive application of point cloud data. The high amount of data collected with LiDAR, RGB have to be transformed into representations that fulfill the computational requirements for VR and AR setups \cite{ARVR}.\\
This paper proposes to use compressed geometric arrays for point cloud representation in various point cloud processes. Our simulation results show that using this representation format, particularly the sparse point cloud data could be represented in a suitable way for computation intensive point cloud processes. This way, using the CGA, the computations complexity of the mentioned applications could be improved.

\section{Compressed Geometric Arrays}
\label{PM}
We propose an efficient array based data representation for 3D point cloud that reduces the amount of stored data and increases the memory bandwidth efficiency. Our proposed approach makes the indirection minimum using the geometric arrays to represent the point cloud data. Since, the indexing in array is much faster and easier. Reducing the number of indirection makes the memory bandwidth utilization much higher.

\begin{figure}[H]
\centering
  \includegraphics[width=0.5\textwidth]{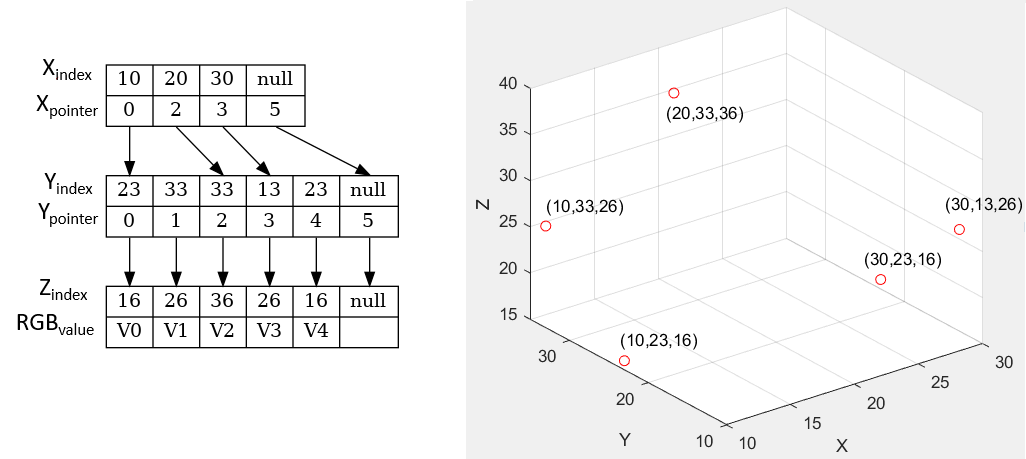}
	\caption{Presentation of six points in a 3D space and the points in the proposed geometric arrays.}
	\label{fig:g-arrays}
\end{figure}

To process a point cloud, the $x$, $y$, and $z$ coordinates of the points should be traversed to find the existing points. For instance, in merging process, in which two octree should be merged, the second octree should be traversed to find each point and then, the first octree should be traveled to find the appropriate place to add this point.
Fig. \ref{fig:g-arrays} shows an example of the CGA data layout for six points (i.e., $\mathbf{p_0}$ to $\mathbf{p_5}$) with different coordinates. We have six arrays in this format.

\begin{itemize}
	\item A $Value$ array stores the attribute values of the existing points in the point cloud.
	\item A $Z_{index}$ vector shows the $z$ coordinates of the corresponding points in the $Value$ array.
	\item A $Y_{index}$ vector shows the $y$ coordinates of the points in the $Value$ array.
	\item A $Y_{pointer}$ vector stores the cumulative number of existing points with the $y$ coordinate equal to $Y_{index}[i]$. It is defined by the following recursive relation.
	\begin{itemize}
		\item $Y_{pointr}[0] = 0$
		\item $Y_{pointer}[i+1] = Y_{pointer}[i]$ + number of existing points with their $y$ coordinates equal to $Y_{index}[i]$
	\end{itemize}
	\item A $X_{index}$ vector shows the $x$ coordinates of the points in the $Y_{index}$ array.
	\item An $X_{pointer}$ vector stores the cumulative number of existing points with the $x$ coordinate equal to the index of $X_{pointer}$ vector. It is defined by the recursive relation below.
	\begin{itemize}
		\item $X_{pointer}[0] = 0$
		\item $X_{pointer}[i+1]$ = $X_{pointer}[i]$ + number of existing points with $x$ coordinate equal to $i$ and non equal $y$ values
	\end{itemize}
\end{itemize}

The proposed data structure is called compressed geometric array, since in this format the duplicated coordinates are not saved as many times as they are repeated. Only the number of times that $x$ and $y$ coordinated are repeated is saved in $X_{pointer}$ and $Y_{pointer}$ arrays, respectively. Therefore, this data structure requires less memory space and is much more compact than when all the coordinates are stored.\\
CGA is an extension of geometric arrays (G-Arrays) for point cloud processing proposed in \cite{G-Arrays-ICASSP2021}. Another array, which is called $X_{index}$, has been added in CGA that shows the $x$ coordinates of the points. Then, the $X_{pointer}$ vector stores the cumulative number of existing points with the $x$ coordinate equal to the index of $X_{pointer}$ vector.
In G-Arrays, where the $X_{index}$ array did not exist, one cell of $X_{pointer}$ array is assigned to each number between the minimum and maximum values of $x$ coordinates. Then, if there is no point in the space with a specific coordinate, the corresponding cell in the $X_{pointer}$ array would be zero. Thus, due to the sparsity of points in 3D point clouds, lots of cells in $X_{pointer}$ array would be zero, which would lead to polluting the memory. In CGA format, just the existing $x$ coordinate are stored in $X_{index}$ array and the corresponding $X_{pointer}$ array would be more compact.\\
In should be noted that the contribution of this paper over our previous work in \cite{G-Arrays-ICASSP2021} is that here, we examine new methods for point cloud processing, including cloud merging, 3D to 2D projection, nearest neighbor (NN) search, and point cloud compression, based on CGA .

\subsection{Point Cloud Merging}
\label{Point cloud merging}
\begin{figure}[H]
\centering
  \includegraphics[width=0.5\textwidth]{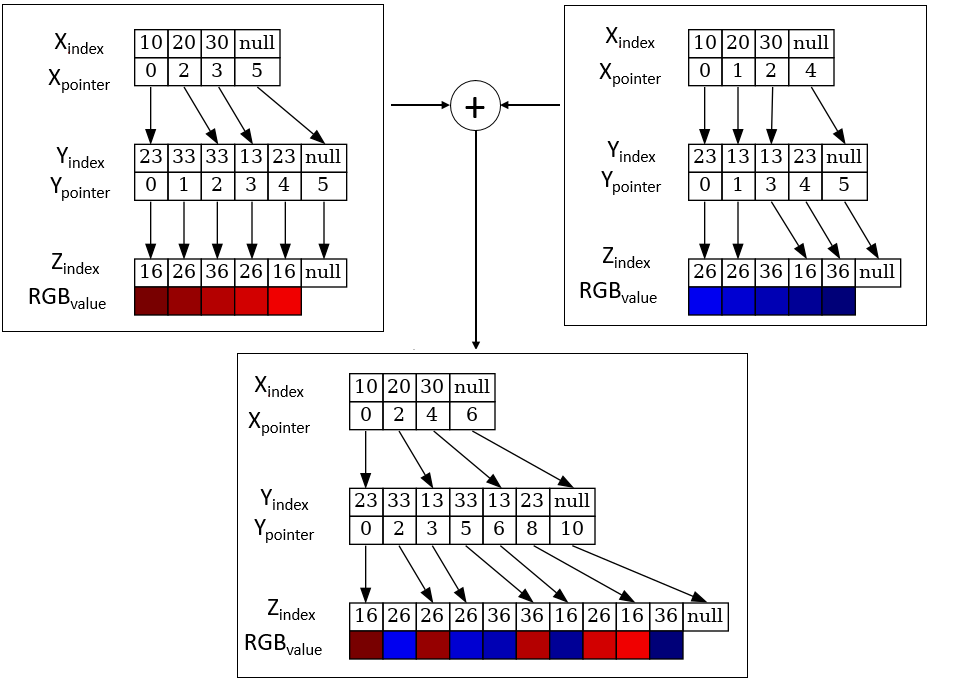}
	\caption{Illustrative example of merging two CGA-based point clouds.}
	\label{g-array_merging}
\end{figure}

Consider merging two point clouds stored in separate CGAs as shown in Fig. \ref{g-array_merging}. To merge these two point clouds, we should concatenate the fields (e.g., dimensions and attributes) of these two different point clouds. One can create a new CGA; then, access the points of each CGA and add them it this new one. To access a specific point $\mathbf{s_0}=(s^x_0, s^y_0 ,s^z_0)$ in CGA, we should first determine if this point exists or not. For this purpose, it is enough to refer to $X_{index}$ array. Then, if $X_{pointer}[s^x_0+1] - X_{pointer}[s^x_0]$ is non-zero, there exist at least one point with the $x$ coordinate equal to $s^x_0$. Then, the pointers of $X_{pointer}[s^x_0]$ and $X_{pointer}[s^x_0+1]$ show the candidate locations for finding the corresponding $y$ coordinates in the  $Y_{index}$ array. Searching the cells with indices between these two pointers in the $Y_{index}$ array, the point with $x$ and $y$ coordinates equal to $s^x_0$ and $s^y_0$ could be found. Then, the index of the point found in  $Y_{pointer}$ array (i.e., $k$) is used to find the corresponding $z$ coordinate. The $Y_{pointer}[k+1] - Y_{pointer}[k]$ shows the number of points with $x$ and $y$ coordinates equal to $s^x_0$ and $s^y_0$, respectively and the pointers of $Y_{pointer}[k]$ and $Y_{pointer}[k+1]$ in $Y_{pointer}$ array show the candidate locations for finding the corresponding $z$ coordinate in the $Z_{index}$ array.

\subsection{Point Cloud 3D to 2D Projection}
Point could projection implements a single view projection of a 3D point cloud that resulting in a 2D plane. For this purpose, each points of the point cloud should be accessed and processed using the proper projection matrix, i.e., orthogonal or perspective projection as explained in ~\ref{Pointcloudprojection}. So, we should access the points stored in CGA as explained in \ref{Point cloud merging}. Then, the orthogonal or perspective projection matrices described in \ref{Pointcloudprojection} are used for each extracted point, $\mathbf{s_0}=(s^x_0, s^y_0 ,s^z_0)$, from the CGA to find the projected point. 

\subsection{Point Cloud NN Search}

We develop a CGA-based NN algorithm for point cloud processing, which finds the nearest point to a given query point for the compression method described in Section \ref{compression_CGA}. To find the nearest point to a query point, we first look for a point with exactly equal coordinates to those of the query point. Using the look up method explained in Section \ref{Point cloud merging}, we determine if such a point exists.
If not, the $X_{index}$ array is searched for the nearest $x$ coordinate to the query point. Then, following the corresponding pointers in $X_{pointer}$ and $Y_{pointer}$ arrays, the nearest $y$ and $z$ coordinates to the query point are identified.
Next, we calculate the Euclidean distance (i.e., $\epsilon$) between the found point and the query point. Any existing points from the set whose $x$, $y$, and $z$ coordinates are distant from the corresponding coordinates of the query point at less than $\epsilon$ are candidates for the nearest point. The candidate points (i.e., $s^x_i$) are identified in $X_{index}$, by calculating the Euclidean distance between each element of $X_{index}$ and the query point.
Note that $X_{pointer}[s^x_i+1] - X_{pointer}[s^x_i]$ shows the number of existing points in the candidates list. Following the corresponding pointers in $X_{pointer}$ and $Y_{pointer}$ arrays, the corresponding $y$ and $z$ coordinates are found. If the distance between the found $y$ and $z$ coordinates from the query point are less than $\epsilon$, these points are candidates for the nearest point. Finally, among these candidate points, a point with the minimum Euclidean distance is selected. To evaluate the proposed NN search algorithm, a bunch of NN queries, using random in the bounding box, are performed.

\subsection{Point Cloud Compression with CGA} 
\label{compression_CGA}
\begin{figure}[H]
\centering
  \includegraphics[width=0.5\textwidth]{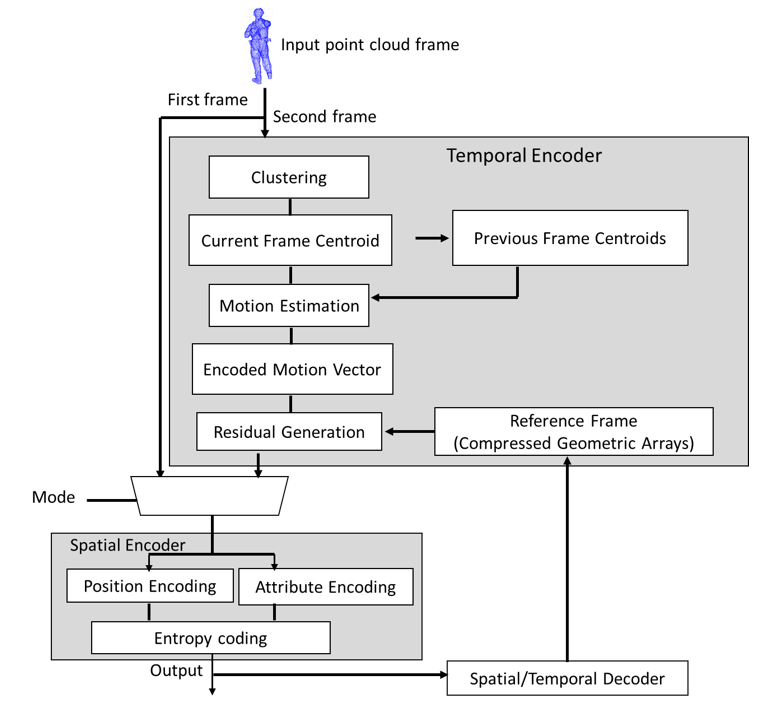}
	\caption{Our proposed point cloud spatio-temporal encoding based on CGA.}
	\label{fig:Proposed Method}
\end{figure}

The proposed method for spatio/temporal compression of point cloud using CGA as presented in Fig. \ref{fig:Proposed Method}. The first frame is coded spatially based on the prior work on MPEG G-VCC \cite{MPEG-G-PCC} and considered as the reference for the second frame. We employ the second path of Fig. \ref{fig:Proposed Method} to consider all the other frames as predicted frames.

\subsubsection {Clustering Points} 
The \textit{clustering} step is proposed to find the motion vectors as the first step in the proposed temporal prediction technique.
Using clustering, parts of the consecutive frame with high similarity are most likely placed in the corresponding clusters of reference and predicted frames. Since, in the motion estimation process, we want to look for the most similar point in the reference frame for each point of the predicted frame, the corresponding cluster of each point in the reference frame is enough to find the most similar point. \\
Without loss of generality, in this paper, the k-means algorithm is used for clustering of the reference and predicted frames. The right number of clusters in k-means algorithm depends on a trade-off between the number of distance computations and the quality of clustering. Based on the minimum and maximum values of the $x$, $y$ and $z$ coordinates, we consider eight initial center points at the corners of each frame . 

\subsubsection {Estimating Cluster Motion}
\begin{figure}[H]
\centering
  \includegraphics[width=0.4\textwidth]{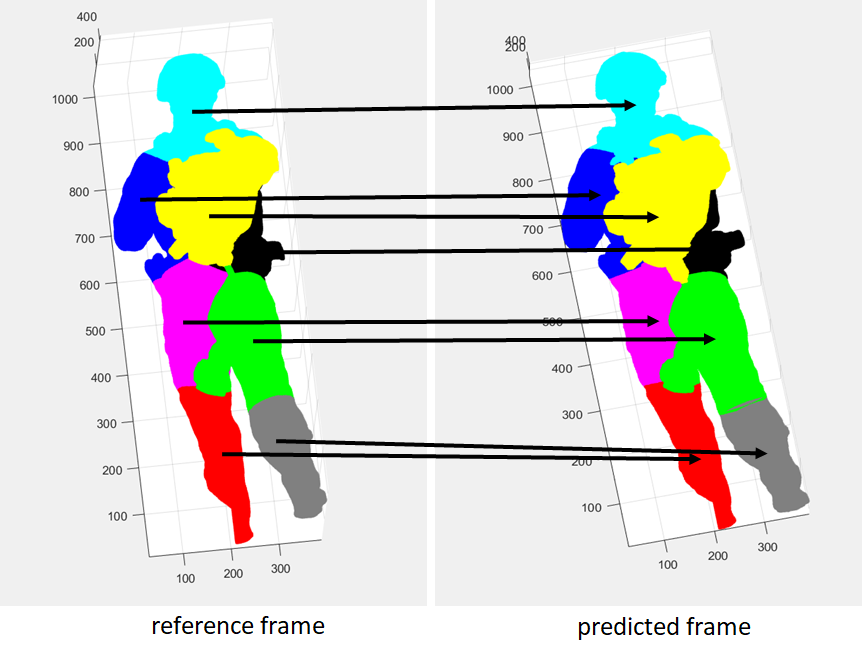}
	\caption{Finding motion vectors based on the clustering of points in the reference and predicted frames.}
	\label{motion_vectors}
\end{figure}

In temporal prediction, the movement of textures or objects in consecutive frames described by Motion estimation process. When the best matching block for a predicted block is found in the reference frame, the corresponding movement between the reference and predicted blocks is represented by motion vector.\\
In the proposed approach, the centroids of the computed clusters in the previous step is used to estimate the coarse-grained motion vectors as shown in Fig. \ref{motion_vectors}. 
The amount of changes in the centroids of the corresponding clusters from the reference and predicted frames could be considered as a good estimation of the amount of movement between the points, since, each cluster contains points with similar coordinates. 
Therefore, the motion vectors are computed using the centroids and are coded as part of the output bitstream.

\subsubsection {Generating Point Residual}
After computing motion vectors, a reconstructed frame  is generated using the predicted frame and the estimated motion vectors. The points of each cluster are displaced according to its corresponding motion vector.\\
The reconstructed error frame, often referred to as the residual frame contains the information to correct the error of motion estimation process. For this purpose, the difference between the attributes of the corresponding points in the reference and reconstructed frames should be computed. But, the main issue is how to find the corresponding point to each point of the reconstructed block in the reference frame. The geometry of points in point clouds, unlike the computer-generated or natural 2D videos, has limited spatio-temporal locality. It means that unlike in 2D video frames, there is no guaranty that all points exist in a 3D point cloud frame. So, finding the corresponding point in reference frame to each specific point in the current frame is not trivial. We examine our proposed CGA format to remove the spatio/temporal redundancy to achieve higher compression ratio. To find the corresponding point to each point of the reconstructed block in the reference frame as the main issue in temporal prediction, our proposed scheme suggests to represent the reference frame in CGA to search its points more quickly. A matching criterion such as Mean Squared Error (MSE) or Sum of Absolute Differences (SAD) can be used to find the corresponding point. SAD is preferable for VLSI implementation because of its simple computational steps. Therefore, we use SAD for choosing the best corresponding block. After searching, the smallest SAD candidate is chosen as the best matching reference for each point of the reconstructed frame. Then, the residual frame is created by calculating the difference between the RGBs of the reconstructed and the reference frame points.

\section{Evaluation}

\subsection{Experimental setup}
To evaluate various point cloud processes, i.e., point cloud merging, 3D to 2D projection, NN search and point cloud compression, a system with Intel(R) Core(TM) i7-8750H CPU @ 2.5GHz 2.6 GHz and 8.00 GB RAM, and Microsoft Windows 10 is used. For the proposed method, the proposed CGA is used to implement the mentioned point cloud processes. As the baseline method, these point cloud processes are implemented using octree data structure of PCL library. PCL splits into a series of modular libraries such as pcl\textunderscore octree library which provides efficient methods for creating a hierarchical tree data structure from point cloud data. This library enables spatial partitioning, downsampling and the other operations on the point data set. Each octree node has either eight children or no children. The root node defines a cubic bounding box that includes all points. At every tree level, this space becomes subdivided by a factor of 2 \cite{Rusu_ICRA2011_PCL}. Since PCL is released under the terms of the BSD license, and free for commercial and research use, we have used the pcl\textunderscore octree library as a basline to implement various point cloud processes.

\subsection{Dataset}
The Moving Picture Expert Group (MPEG) has suggested a coding solution for various categories of point clouds: 1) LIDAR point cloud for dynamically acquired data, 2) Surface point cloud for static point cloud data, and 3) Video-based point cloud for dynamic content. According to these categories, Video-based point cloud processing (V-PCC) and Geometry-based point cloud processing (G-PCC) standards have been developed \cite{MPEG}, \cite{VPCC_MPEG} and \cite{V_G_PCC}. Video-based point cloud focused on point sets with a relatively uniform distribution of points and LIDAR and surface point clouds are appropriate for more sparse distributions. 
In this paper, LIDAR and surface point cloud and the corresponding video coding standard (G-PCC) are used in our experiments. As surface point clouds, the 8i Voxelized Surface Light Field (8iVSLF) Dataset \cite{8iVSLF} is used. For each point cloud in the 8iVSLF dataset, the full body of a human subject was captured by 39 synchronized RGB cameras, at 30 frames per second (fps). In our experimental results, Soldier, Longdress, RedandBlack, and Loot are from 8iVSLF dataset. The other two point cloud sequences, GT\textunderscore Madame and Urbane scene \cite{Paris-Lille-3} are LIDAR point cloud. GT\textunderscore Madame contains two PLY files, each one with 10 million points that are collected from rue Madame, a street in France. The Urbane scene dataset consists of around 2km of Mobile Laser Scanning point cloud acquired in two cities. Each point cloud consists of a list of $x$, $y$ and $z$ corresponding to geometric coordinates and the corresponding attributes. 

\subsection{Simulation Results}
In this section, we assess the performance of CGA for point cloud processing, i.e., merging, projection, NN search and compression of 3D point clouds. The point cloud operations are implemented as explained in Section \ref{PM}.
For the baseline method, we have considered the octree structure from PCL library \cite{Rusu_ICRA2011_PCL}.\\
First, we compare the required time for octree and CGA construction for various point cloud sequences as shown in table ~\ref{tab1}. The results show that in terms of the construction time, CGA is superior to the octree data structure.

\begin{table}[htb]
	\caption{Comparison of the required time for octree and CGA construction}
	\begin{center}
		\begin{tabular}{ccc}
			\hline
			Point cloud sequences & \multicolumn{2}{c}{The required time for generating data structure} \\
			\hline			
			& Octree & CGA \\
			\hline
			Soldier & 3.93s & 3.87s\\
			\hline
			Longdress & 2.68s & 1.90s \\
			\hline
			Loot & 2.94s & 2.49s \\
			\hline
			RedandBlack & 2.75s & 2.28s\\
			\hline
		\end{tabular}
		\label{tab1}
	\end{center}
\end{table}

\begin{figure*}[htb!]
\centering
  \includegraphics [width=0.95\textwidth,  height=0.95\textheight]{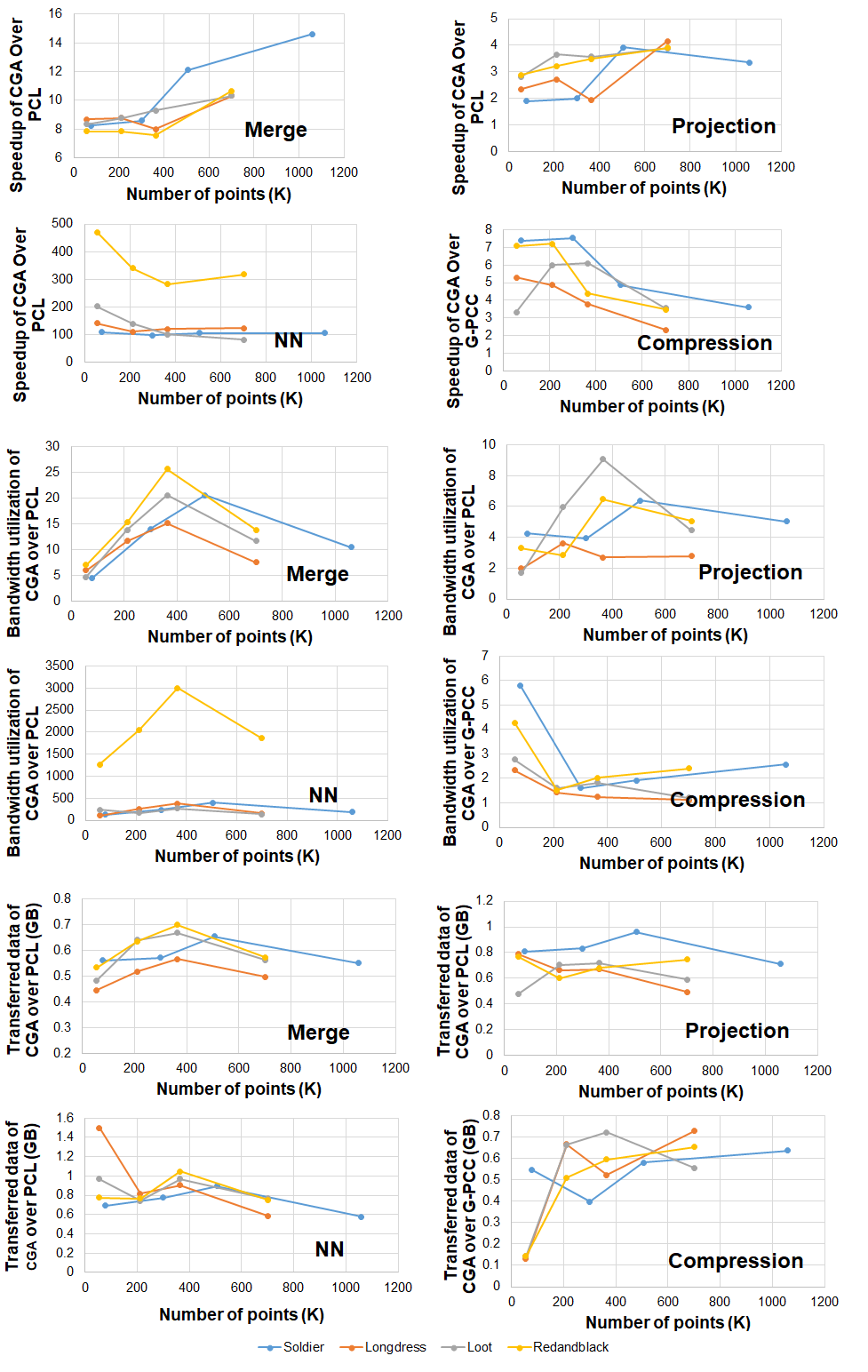}
	\caption{Speedup, memory bandwidth utilization, and the volume of transferred data of the CGA over PCL for computer-generated point cloud sequences.}
	\label{G-arrays_PT}
\end{figure*}

\begin{figure*}[htb!]
\centering
  \includegraphics  [width=0.95\textwidth,  height=0.95\textheight]{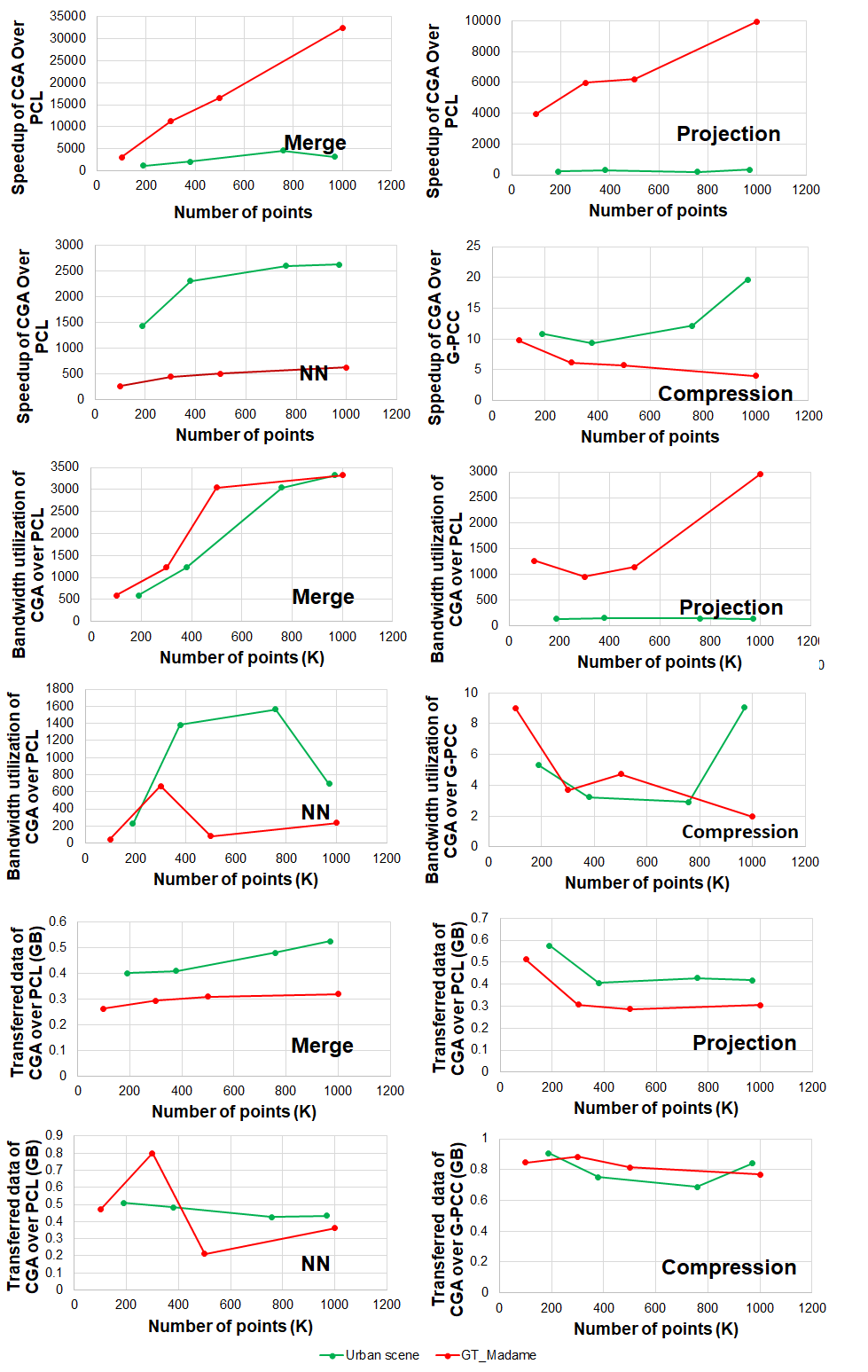}
	\caption{Speedup, memory bandwidth utilization, and the volume of transferred data of the CGA over PCL for LiDAR point cloud sequences.}
	\label{G-arrays_LiDAR}
\end{figure*}

We measure performance in terms of the processing time, memory bandwidth, and the amount of transferred data between CPU and memory.
We compute the speed up, memory bandwidth, and the amount of transferred data of CGA over PCL for the computer-generated and LiDAR sequences as shown in Fig. \ref{G-arrays_PT} and Fig. \ref{G-arrays_LiDAR}, respectively. The x-axis shows the number of points for each sequence. The results indicate that CGA can replace octrees in point cloud processing to improve the processing time significantly. \\

The reason for the significant improvement in the processing time of LIDAR point cloud compared to the computer-generated ones is that the LIDAR point cloud data used in this paper has no color attributes and the geometric values are much smaller than the computer-generated point cloud sequences. So, the amount of data that should be processed is much less than the computer-generated ones. Table ~\ref{tab2} shows six points from computer-generated and LIDAR point cloud sequences for more clarifications.

\begin{table}[htb]
	\caption{The $x$, $y$ and $z$ coordinates and color attributes of computer-generated and LIDAR point cloud sequences}
	\begin{center}
		\begin{tabular}{cccccc|ccc}
			\hline
			\multicolumn{6}{c|}{Soldier (computer-generated)} & \multicolumn{3}{c}{GT\textunderscore Madame (LIDAR)} \\
			\hline			
			 $x$ & $y$ & $z$ & red & green& blue	& $x$ & $y$ & $z$\\
			\hline
			127 & 15 & 157 & 82 & 73 & 67 & 19 & 19 & 19 \\
			\hline 
			127 & 14 & 159 & 82 & 73 & 67 & 7 & 7 & 7 \\
			\hline
			127 &15 & 158 & 82 & 73 & 67 & 17 & 17 & 17 \\
			\hline
			127 & 15 & 159 & 82 & 73 & 67 & 18 & 18 & 18 \\
			\hline
			127 & 17 & 155 & 79 & 72 &67 & 11 & 11 & 11 \\
			\hline
			127 & 19 & 153 & 83 & 77 & 73 & 15 & 15 & 15 \\
			\hline
		\end{tabular}
		\label{tab2}
	\end{center}
\end{table}

The normalized bandwidth utilization of CGA over PCL for the computer-generated and LiDAR sequences, showed in Fig. \ref{G-arrays_PT} and Fig. \ref{G-arrays_LiDAR}, demonstrate the effectiveness of the proposed CGA approach to achieve much higher bandwidth utilization compared to PCL.\\
The existence of peaks and valleys depends on the type of point cloud data and the hardware configuration. The simulations are done on real hardware and the number of cache misses affects the bandwidth utilization considerably.
As explained in section \ref{Performance_analysis}, the number of points are changed by downsizing the point cloud to three different versions. Then, the simulations are done for these downsized versions and the original point cloud sequences. In the original point cloud sequences with the maximum number of points, there is a high correlation between the coordinates and the color attributes of the adjacent points. This high correlation between the coordinates and the color attributes of the points greatly reduces the number of cache misses during point cloud processes.\\
Then, we have used the downsampling process to reduced the number of points of each cloud to half of the initial points. Due to the cache size of the system (384 KB), significant volumes of points do not fit in the cache. In addition, the correlation between the coordinates and the color attributes of the points is reduced considerably during downsampling process. So, the number of cache misses during the point cloud operations increase significantly. The peaks in Fig. \ref{G-arrays_PT} and Fig. \ref{G-arrays_LiDAR}, show the related bandwidth utilization for these downsized versions of point clouds.\\
For the other downsized versions of point clouds, the valleys shown in Fig. \ref{G-arrays_PT} and Fig. \ref{G-arrays_LiDAR}, the total number of generated points is so small that most of the points can be fitted in the cache. So, the number of cache misses is reduced considerably during the point cloud operations.\\
One of the main reasons for the superior performance of CGA over PCL is the reduced data transfer during point cloud processing. The amount of transferred data by CGA normalized to that of PCL for the computer-generated and LiDAR sequences is shown in Fig.~\ref{G-arrays_PT} and Fig.~\ref{G-arrays_LiDAR}, respectively. The amount of transferred data in CGA compared to the PCL is improved significantly.\\
We also use CGA for temporal encoding of point clouds. In our proposed approach, the frames of the sequences are coded sequentially. 
Advanced compression methods for point clouds try to remove the spatial and temporal redundancy within a point cloud. For the temporal encoding of a frame (predicted frame), instead of directly encoding the raw pixel values, the encoder finds similar points to the points of the predicted frame from a previously encoded frame (reference frame). Temporal prediction relies on estimating motion between blocks of the consecutive frames. In this process, given a specific block A in the predicted frame and its matching block B in the reference frame, a motion vector (MV) is defined as a vector connecting the top-left coordinates of A and B. Then, the extracted motion vector and a residual block are encoded and sent to the decoder. The residual block is defined as the difference between A and B. Then, the reconstructed error frame, often referred to as the residual frame, should be found. This frame contains the information to correct the error of motion estimation process. For this purpose, the difference between the attributes of the corresponding points in the reference and reconstructed frames should be computed. But, the main issue in point cloud compression is how to find the corresponding point to each point of the reconstructed block in the reference frame. The geometry of points in point clouds, unlike the natural 2D videos, has limited spatio-temporal locality. In other words, there is no guaranty that all points exist in a 3D point cloud frame. So, finding the corresponding point in the reference frame to each specific point in the current frame is not trivial. We need to store the points in a searchable data structure. The data structure may be a sparse matrix because the point cloud doesn’t necessarily include all possible points in the 3D space. Storing the non-existing points would lead to polluting the memory. Even in its sparse representation form, a typical point cloud comprises millions of points, which imposes significant pressure on the memory capacity and bandwidth.\\
Our proposed CGA format enables a faster lookup than octrees to remove the spatio/temporal redundancy during the compression process. The results in table ~\ref{tab3} indicate that CGA can replace octrees in point cloud compression to enable significantly faster point lookups. For the baseline method, all of the frames are coded using their best set of parameters in the MPEG G-PCC open-source library \cite {b1}.

\begin{table}[htb]
	\caption{Performance gains of the CGA over the octree in compression process}
	\begin{center}
		\begin{tabular}{cc}
			\hline
			Point cloud sequences	 & Lookup speedup during the compression process\\
			&  over octree \\
			\hline
			Soldier & 16.59x \\
			\hline
			Longdress & 15.97x \\
			\hline
			RedandBlack & 23.58x \\
			\hline
			Loot	& 54.78x \\
			\hline
		\end{tabular}
		\label{tab3}
	\end{center}
\end{table}

\begin{figure*}[htb!]
\centering
  \includegraphics [width=0.8\textwidth]{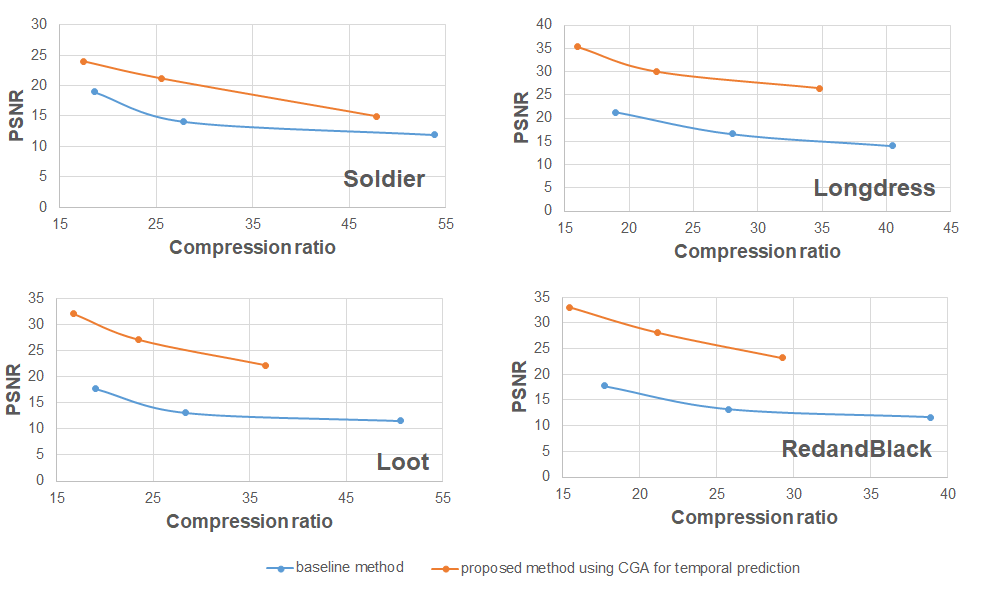}
	\caption{The PSNR versus compression ratio for various quantization parameters (QP) for baseline method and our compression scheme using CGA.}
	\label{PCR}
\end{figure*}

We also measure performance in terms of the output quality and compression ratio. There are several objective quality metrics available in the literature to measure the quality of encoded point clouds.
The point-to-point metric measures the quality using the point-based distances. For each point in the decoded frame, the nearest neighbor in the original point
cloud frame is obtained and a distance-based mean squared error (MSE) is computed over all pairs of points. The point-to-plane metric is often used to better evaluate the output quality, since, the point cloud points represent the surface of objects in a visual scene. This metric approximates the underlying surface at each point as a plane. This metric results in smaller errors for the points that are closer to the surface \cite{distortion_metrics}. We study various quantization parameters (QP = 11, 10, and 9) for our simulations which determines the number of bits representing each component of the points.  Fig. \ref{PCR} shows the comparison of rate-distortion of the baseline and our proposed method.

\section{Analysis of the time and space complexity}

For various point cloud operations, the constructed point cloud should be traveresd to access the points.
For octree construction, a three-dimensional space between the minimum and maximum coordinates of points is recursively subdividing into eight octants. Each octant is further subdivided recursively only when they contain some points. The final subdivision results in eight leaf nodes that store points. In each division, the three-dimensional space is divided by two in each dimension. So, the maximum depth of the octree would be log(N), where the space is of dimension N x M x L and N $\geq$ M, L.
Hence, using the octree structure, accessing the points of a point cloud would be done in O(KlogN), where the space is of dimension N x M x L, N $\geq$ M, L, and K is the number of points. \\
But in CGA, to access the points of a point cloud, the $X_{index}$ array should be traversed to access the $x$ coordinates of the points. Then, following the corresponding pointers in $X_{pointer}$ and $Y_{pointer}$ arrays of each $x$ coordinate, the nearest  $y$ and $z$ coordinates to the query point are identified in O(1). So, the time complexity of accessing a point in CGA is O(K), where K is the number of points. This way, the time complexity of using CGA in various point cloud processes are much less than octree structure.\\
The only avaiable data structure in the litrature that uses pointers similar to our proposed CGA is windowed priority queue (WinPQ) \cite{WinPQ} that holds references to points, sorted in one dimension. Intervals of points are extracted from the WinPQ. Each interval contains all points within a specific one-dimensional window. The interval of points can be updated by moving the window in discrete steps. The one-dimensional WinPQ can be used to scan higher-dimensional point cloud data by instantiating it repeatedly within the nested loops. So, for three-dimension point cloud the time complexity of point cloud operations using WinPQ is O(K$^{3}$), where K is the number of points. So, WinPQ is much more complex compared to our proposed CGA.\\
We also evaluate the performance of the CGA as compared to octrees in terms of memory consumption. For this purpose, an octree and a CGA are constructed to store the points of computer-generated point clouds. Table ~\ref{tab5} shows the memory saving of CGA structure over the Octree.

\begin{table}[htb]
	\caption{Memory saving of CGA structure over the Octree}
	\begin{center}
		\begin{tabular}{cc}
			\hline
			Sequence name	& Memory saving of CGA structure\\
			& over octree \\
			\hline 
			Soldier & 1.44x \\
			\hline
			Longdress & 1.43x \\
			\hline
			RedandBlack & 1.42x \\
			\hline
			Loot & 1.40x \\
			\hline
		\end{tabular}
		\label{tab5}
	\end{center}
\end{table}

\section{Conclusion}
In this paper, we proposed a new data structure for processing point clouds, called CGA. We examined various point cloud operations, including merge, projection, NN search and point cloud compression. Our simulation results on a set of computer-generated and LiDAR point cloud sequences showed that the proposed format has provided significant speed ups, better bandwidth utilization, and less transferred data compared to the state-of-the-art tree-based structures from the PCL library.


%
%



\end{document}